# Aggregate Hardware Impairments Over Mixed RF/FSO Relaying Systems With Outdated CSI

Elyes Balti, *Student Member, IEEE*, Mohsen Guizani, *Fellow, IEEE*,
Bechir Hamdaoui, *Senior Member, IEEE*, and Bassem Khalfi, *Student Member, IEEE*

*Abstract*—In this paper, we propose a dual-hop radio-frequency (RF)/free-space optical system with multiple relays employing the decode-and-forward and amplify-and-forward with a fixed gain relaying scheme. The RF channels are subject to a Rayleigh distribution while the optical links experience a unified fading model emcopassing the atmospheric turbulence that follows the Málaga distribution (or also called the $\mathcal{M}$-distribution), the atmospheric path loss, and the pointing error. Partial relay selection with outdated channel state information is proposed to select the candidate relay to forward the signal to the destination. At the reception, the detection of the signal can be achieved following either heterodyne or intensity modulation and direct detection. Many previous attempts neglected the impact of the hardware impairments and assumed ideal hardware. This assumption makes sense for low data rate systems but it would no longer be valid for high data rate systems. In this paper, we propose a general model of hardware impairment to get insight into quantifying its effects on the system performance. We will demonstrate that the hardware impairments have small impact on the system performance for low signal-to-noise ratio (SNR), but it can be destructive at high SNR values. Furthermore, analytical expressions and upper bounds are derived for the outage probability and ergodic capacity while the symbol error probability is obtained through the numerical integration method. Capitalizing on these metrics, we also derive the high SNR asymptotes to get valuable insight into the system gains, such as the diversity and the coding gains. Finally, analytical and numerical results are presented and validated by the Monte Carlo simulation.

## I. INTRODUCTION

WIRELESS optical communications through the atmosphere also known as Free-Space Optical (FSO) communication have recently attracted considerable attention for a wide variety of applications such as fiber backup, back-haul for wireless, cellular networks, redundant links and disaster recovery [1], since it offers cost effective, high bandwidth and licence free access to the optical spectrum. These benefits make the FSO technology as a complementary or an alternative to the RF communication in some applications since it has solved many classical problems that the RF suffers from, e.g., FSO offers full duplex Gigabit Ethernet throughput, immunity to interferences and high security [2].

### A. Motivation

Although FSO links have recently gained enormous attention due to the aforementioned advantages, they are extremely sensitive to atmospheric weather conditions, e.g., rain, fog, snow [3], [4]. Standard values of the atmospheric path loss relative to the weather conditions are detailed and given by [5] and [6]. In addition, FSO communications also suffer from degradations caused by atmospheric turbulences. The origins of these turbulences are the variations of the refractive index of the medium due to the heterogeneity in the temperature and the atmospheric pressure changes. As a result, rapid fluctuations affect the transmitted optical signal which, is referred to turbulence-induced fading or scintillation. Many attempts in the literature considered optical channels under different models characterizing the scintillation fading. The most common used models are Log-Normal [5], [7] and Gamma-Gamma fadings [8], [9]. The first model is widely used for weak atmospheric tubulence conditions. However, the Log-Normal model is not accurate enough to model both the moderate and the strong scintillations. In addition, the probability density function (PDF) of the Log-Normal fading significantly deviates from the experimental data at the tails. Gamma-Gamma fading has recently gained attention, in particular for the work directed to Mixed RF/FSO systems [10], [11], since it is more precise than Log-Normal in modeling the moderate and strong atmospheric tubulences. However, the PDF of Gamma-Gamma suffers from inaccuracy in the tails compared to the experimental data. Since the calculation of the fade and the detection probabilities is essentially based on the tail of the PDF, overestimation or underestimation of the tail region substantially affects the accuracy of the performance analysis and certainly leads to flaw outcome results. To solve this problem, a new efficient optical fading model called Málaga or also called the $\mathcal{M}$-distribution is proposed to consider weak, moderate and strong turbulences modeling. In addition, $\mathcal{M}$-distribution reflects wide ranges of turbulence models such as Log-Normal, Gamma-Gamma, Shadowed-Rice [12], Rice-Nakagami, etc. [13], [14]. Furthermore, it has been proved that the PDF of Málaga fading is more precise and abide to the simulation data than Gamma-Gamma fading in particular at the tail region for the plane wave as well as the spherical wave.

In addition to the path loss and turbulence fading, FSO links are also subject to the pointing error which, is originated from the misalignment between the laser-emitting transceiver and the photodetector. Various factors such as seismic activities, building sways results in the aforementioned misalignment. To characterize this degradation, Uysal *et al.* [15] discuss various models for the radial displacement of the pointing error for a Gaussian laser beam. The most generalized model that covers various special cases is the so-called Beckmann pointing error model. Based on this model, related work has adopted the following choices to model the radial displacement such as Rician [16], Hoyt [17], NonZero-Mean and Zero-Mean Single-Sided Gaussian [18] but the most widely used is Rayleigh [19]–[21] for reason of simplicity.

To achieve better performance, it is important to focus on the optical detection that can be either IM/DD or heterodyne [9], [22], [23]. IM/DD has been recently used as the main detection method. Despite the complexity of the heterodyne implementation at the receiver, it has been shown that it outperforms the IM/DD in combating the atmospheric fading effect [10], [24].

### B. Related Work

Cooperative relaying-assisted communication has recently gained an enormous interest since it offers not only high Quality of Service (QoS), reliability and power-efficient coverage but also improves the capacity of wireless communication systems. Recently, considerable efforts have been conducted to study the asymmetric relaying RF/FSO system under the assumptions of Decode-and-Forward (DF) [25], [26], Amplify-and-Forward (AF) [27], [28], Quantify-and-Encode (QE) [29], [30].

Unlike the opportunistic and best relay selections which, require the full knowledge of the CSI of all links, Krikidis *et al.* [31] proposed a partial relay selection (PRS) which, requires the CSI of only one hop (source-relay or relay-destination). Consequently, the power efficiency is achieved and additional network delay will be avoided. Further details of this protocol will be given later. For time-varying channels, the instantaneous CSI used to choose the relay can be outdated due to the slow propagation of the feedbacks from the relays to the source, so it is certain that the selected relay is not necessarily the best one. This outdated CSI, however, can lead to deteriorate the performance results that highly depends on the time correlation coefficient between the outdated and the actual CSI. Contrary to [31] and [32] where they considered PRS with perfect CSI estimation, outdated CSI of Rayleigh and Nakagami-m fading is assumed in [11], [33], and [34].

### C. Hardware Impairments Overview

Although there are plenty of contributions in mixed RF/FSO systems in the literature [35]–[40], many of them have neglected the effects of hardware impairments. In practice, due to its low quality, hardware (source, relays …) suffers from the many types of impairments, e.g., non-linear High Power Amplifier (HPA) [41], [42] as well as phase noise [43] and I/Q imbalance [44]. It has been shown in [45] that the I/Q imbalance switches the phase of the signal constellation as well as it mitigates its amplitude. Moreover Dardari *et al.* [41] have characterized the impact of HPA non-linearities and concluded that this impairment model creates a distortion of the signal as well as noise. There are many types of the HPA non-linearities impairment model, e.g., Ideal Soft-limiter Amplifier [46], Solid state power amplifier (SSPA), Soft envelope limiter (SEL), Traveling wave tube amplifier (TWTA) [47], [48]. Maletić *et al.* [48] turned out that the TWTA has a severe impact on the system performance in terms of OP and EC than the SEL impairment model. In addition, [41], [45], and [48] demonstrated that the joint impact of the I/Q imbalance and non-linear HPA saturates the system capacity by creating a ceiling that inhibits the capacity from indefinite increase. As we mentioned earlier, the hardware impairments can be negligeable for low-rate systems, while it is more pronounced in the range of high SNR. So to get full insight into our system characteristics and obtain flawless performance results, the hardware impairments should be taken into account.

### D. Contribution

In this work, our contribution is to consider the Double-Weibull fading as a model for the optical channels which, is more accurate than the most common used Log-Normal and Gamma-Gamma distributions. We assume DF and AF with fixed gain relaying due to its low complexity/cost systems, where the low latency originating from the signal processing is of high importance. Moreover, the PRS protocol with outdated CSI is considered in our system to reduce the power consumption dedicated to the relay selection. Furthermore, the most important issue in this work is to introduce an aggregate model of the hardware impairments in the source and the relays as the work done by Bjornson *et al.* [49]. Although, they quantified the impact of the hardware impairments on their system and they obtained good results, they employed a single relay dual-hop RF system. Even though Balti *et al.* [50], [51] proposed a mixed RF/FSO system with general model of hardware impairments, the author considered Gamma-Gamma fading as a model of the optical channels.

In addition, Subcarrier Intensity Modulation (SIM) is implemented into the relays to modulate the intensity of the FSO carriers. Various binary modulation schemes are assumed to validate the error performance of the proposed system. To the best of our knowledge, we are the first group to propose a general model of hardware impairments to a mixed RF/FSO system with multiple relays and assuming the unified $\mathcal{M}$-distribution to model the optical channels. The analysis of this paper follows these steps:

1) Present a detailed analyis of the system and channels' models.
2) Provide the Cumulative Distribution Function (CDF) and the PDF of the RF and FSO channels.

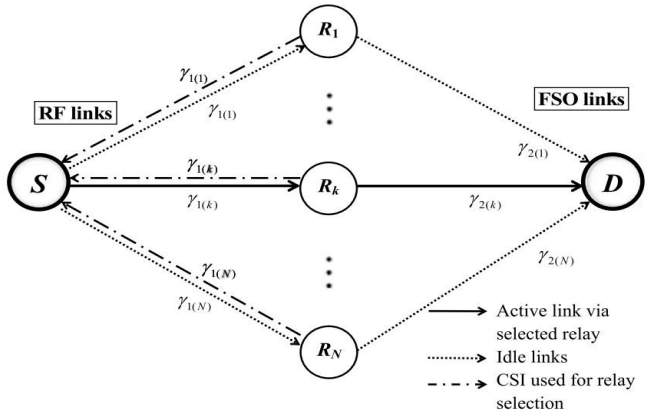

Fig. 1. Mixed RF/FSO system with PRS.

3) Derive the expressions of the end-to-end Signal-to-Noise-plus-Distortion Ratio (SNDR), which is a measure of the degradation of the signal by unwanted or extraneous signals including noise and distortion, for AF and DF relaying protocols.

4) Based on the aforementioned expressions, novel closed-forms, upper bounds as well as high SNR asymptotes of the outage probability (OP), the symbol error probability (SEP) and the ergodic capacity (EC) are derived.

5) Capitalizing on the high SNR asymptotes, engineering insights into the system performance such as the diversity and the coding gains are derived.

### E. Structure

The rest of this paper is organized as follows: Section II describes the system and the channels' models while the outage probability analysis is given in Section III. The symbol error probability and the ergodic capacity analysis are provided in Sections IV and V, respectively. Analytical and simulation results following their discussions are reported in Section VI. Finally, concluding remarks are presented in Section VII.

### F. Notation

For the sake of organization, we provide some useful notations for clarity. Let $f_h(\cdot)$ and $F_h(\cdot)$ denote the PDF and CDF of the random variable $h$, respectively. The Gamma distribution with parameters $\alpha$ and $\beta$ is denoted by $\mathcal{G}(\alpha, \beta)$ while the Shadowed-Rician distribution with parameters $b_0$, $\beta$, $\rho$, $\Omega$ and $\Delta \phi$ is given by $\mathcal{SR}(b_0, \beta, \rho, \Omega, \Delta \Phi)$. In addition, the Gaussian distribution of parameter $\mu$, $\sigma^2$ is denoted by $\mathcal{N}(\mu, \sigma^2)$. The operator $\mathbb{E}[\cdot]$ stands for the expectation while $\mathrm{Pr}(\cdot)$ denotes the probability measure. The symbol $\curvearrowright$ stands for "distributed as".

## II. System Model

Our system consists of a source (S), a destination (D) and $M$ parallel relays wirelessly connected to (S) and (D) as shown in Fig. (1). To select the relay of rank $m$, we refer to the PRS with outdated CSI to select the best one primarily on the feedback of the CSI coming from the relays. For a given transmission, (S) receives the feedbacks about the CSI of the RF channels from the $M$ relays ($\tilde{\gamma}_{1(l)}$ for $l = 1,...M$) and arranges them in an increasing order of amplitudes as follows: $\tilde{\gamma}_{1(1)} \leq \tilde{\gamma}_{1(2)} \leq \ldots \leq \tilde{\gamma}_{1(M)}$. The perfect scenario is to pick the best relay ($m = M$). However, the best relay is not always available due to its half-duplex operation mode. In this case, (S) will choose the next best available relay. Consequently, the PRS consists of choosing the $m$th worst or ($M-m$)th best relay $R_m$. Taking into account the feedback delays as well as the time-varying channels, the CSI used for the relay selection is different from the actual CSI used for transmission. Hence, the outdated CSI should be assumed instead of the perfect CSI estimation. As a consequence, the instantaneous SNRs used for relay selection and transmission are correlated with time correlation coefficient $\rho_m$.

The received signal at the $m$th relay is given by:

$$y_{1(m)} = h_m(s + \eta_1) + \nu_1, \tag{1}$$

where $h_m$ is the fading amplitude of the RF channel between (S) and $R_m$, $s \in \mathbb{C}$ is the information signal, $\nu_1 \curvearrowright \mathcal{N}(0, \sigma_0^2)$ is the AWGN of the RF channel, $\eta_1 \curvearrowright \mathcal{N}(0, \kappa_1^2 P_1)$ is the distortion noise at the source (S), $\kappa_1$ is the impairment level in (S) and $P_1$ is the average transmitted power from (S).

### A. End-to-End SNDR: Amplify-and-Forward Relaying

After reception, the relay $R_m$ amplifies the received signal $y_{1(m)}$ with a fixed gain $G$ depending on the average electrical channel fading. The gain $G$ can be defined as follows [49, eq. (11)]:

$$G^2 \triangleq \frac{P_2}{P_1 \mathbb{E}\left[|h_m|^2\right](1 + \kappa_1^2) + \sigma_0^2}, \tag{2}$$

where $P_2$ is the average transmitted power from the relay to the destination (D). Then, the relay $R_m$ converts the electrical signal to the optical one which, is defined as follows:

$$y_{opt(m)} = G(1 + \eta_e)y_{1(m)}, \tag{3}$$

where $\eta_e$ is the electrical-to-optical conversion coefficient.

At the destination, the signal can be detected following the IM/DD or the heterodyne detection. The signal at the node (D) can be written as follows:

$$y_{2(m)} = (\eta_o I_m)^{\frac{r}{2}}[G(1 + \eta_e)(h_m(s + \eta_1) + \nu_1) + \eta_2] + \nu_2, \tag{4}$$

where $\eta_o$ is the optical-to-electrical conversion, $I_m$ is the optical channel gain between $R_m$ and D, $\eta_2 \curvearrowright \mathcal{N}(0, \kappa_2^2 P_2)$ is the distortion noise at the relay $R_m$, $\kappa_2$ is the impairment level in $R_m$, $\nu_2 \curvearrowright \mathcal{N}(0, \sigma_0^2)$ is the AWGN of the optical channel.

The SNDR depends on the instantaneous electrical $\gamma_{1(m)}$ and the optical $\gamma_{2(m)}$ SNRs of the two hops. They are respectively defined by:

$$\gamma_{1(m)} = \frac{|h_m|^2 P_1}{\sigma_0^2} = |h_m|^2 \mu_1, \tag{5}$$

where $\mu_1 = \frac{P_1}{\sigma_0^2}$ is the average SNR of the first hop.

The parameter $r$ takes two values 1 and 2 standing for heterodyne and IM/DD, respectively. The average SNR $\overline{\gamma}_r$ and the average electrical SNR $\mu_r$ can be expressed as follows:

$$\overline{\gamma}_r = \frac{\mathbb{E}\left[I_m^r\right]}{\mathbb{E}\left[I_m\right]^r}\mu_r, \tag{6}$$

$$\mu_r = \frac{\eta^r \mathbb{E}\left[I_m\right]^r}{\sigma_0^2}, \tag{7}$$

Hence, the instantaneous optical SNR $\gamma_{2(m)}$ can be obtained by:

$$\gamma_{2(m)} = \frac{(\eta_o I_m)^r}{\sigma_0^2}, \tag{8}$$

For ideal hardware, the end-to-end SNR is given by [11, eq. (6)]:

$$\gamma_{\text{id}} = \frac{\gamma_{1(m)}\gamma_{2(m)}}{\gamma_{2(m)} + C}, \tag{9}$$

In case of non-ideal hardware and after some mathematical manipulations, the end-to-end instantaneous SNDR can be expressed as follows [49, eq. (13)]:

$$\gamma_{\text{ni}} = \frac{\gamma_{1(m)}\gamma_{2(m)}}{\delta\gamma_{1(m)}\gamma_{2(m)} + (1+\kappa_2^2)\gamma_{2(m)} + C}, \tag{10}$$

where $\delta \triangleq \kappa_1^2 + \kappa_2^2 + \kappa_1^2\kappa_2^2$ and $C = \mathbb{E}\left[\gamma_{1(m)}\right](1+\kappa_1^2)+1$.

### B. End-to-End SNDR: Decode-and-Forward Relaying

In case of DF relaying protocol, the signal is transmitted only if the relay is able to decode it. Therefore, the effective SNDR is the minimum of the set of SNDRs between (source-relay) and (relay-destination). In case of ideal hardware ($\kappa_1 = \kappa_2 = 0$), the end-to-end SNR is expressed as follows [49, eq. (18)]:

$$\gamma_{\text{id}} = \min(\gamma_{1(m)}, \gamma_{2(m)}), \tag{11}$$

In case of non-ideal hardware, the end-to-end SNDR is written as follows [49, eq. (17)]:

$$\gamma_{\text{ni}} = \min\left(\frac{\gamma_{1(m)}}{\kappa_1^2\gamma_{1(m)} + 1}, \frac{\gamma_{2(m)}}{\kappa_2^2\gamma_{2(m)} + 1}\right), \tag{12}$$

### C. Channel Model

*1) Statistics of the RF Channel:* Since the channel gain $h_m$ experiences a Rayleigh fading, the instantaneous SNR $\gamma_{1(m)}$ is exponentially distributed. Given that PRS with outdated CSIs is assumed, the CDF of $\gamma_{1(m)}$ can be obtained by [52, eq. (9)]:

$$f_{\gamma_{1(m)}}(\gamma) = m\binom{M}{m}\sum_{k=0}^{m-1}\frac{(-1)^k\binom{m-1}{k}}{[(M-m+k)(1-\rho_m)+1]\mu_1}$$
$$\times \exp\left(-\frac{(M-m+k+1)\gamma}{[(M-m+k)(1-\rho_m)+1]\mu_1}\right), \tag{13}$$

After some mathematical manipulations, the CDF of $\gamma_{1(m)}$ can be expressed as follows:

$$F_{\gamma_{1(m)}}(\gamma) = 1 - m\binom{M}{m}\sum_{k=0}^{m-1}\binom{m-1}{k}\frac{(-1)^k}{M-m+k+1}$$
$$\times \exp\left(-\frac{(M-m+k+1)\gamma}{[(M-m+k)(1-\rho_m)+1]\mu_1}\right), \tag{14}$$

TABLE I
PARAMETERS OF THE FSO PART

| Parameter | Definition |
|---|---|
| $\sigma$ | Weather attenuation |
| $\sigma_s^2$ | Jitter variance |
| $\sigma_R^2$ | Rytov variance |
| $k$ | Wave number |
| $\lambda$ | Wavelength |
| $\xi$ | Pointing error coefficient |
| $\omega_0$ | Beam waist at the relay |
| $\omega_L$ | Beam waist |
| $\omega_{\text{Leq}}$ | Equivalent beam waist |
| $L$ | Length of the optical link |
| $a$ | Radius of the receiver aperture |
| $A_0$ | Fraction of the collected power at L = 0 |
| $F_0$ | Radius of curvature |
| $C_n^2$ | Refractive index of the medium |
| $R$ | Radial displacement of the beam at the receiver |

The constant $C$ mentioned earlier depends on the expression of $\mathbb{E}\left[\gamma_{1(m)}\right]$ which, can be obtained by [53, eq. (6)]:

$$\mathbb{E}\left[\gamma_{1(m)}\right] = m\binom{M}{m}\sum_{k=0}^{m-1}\binom{m-1}{k}(-1)^k$$
$$\times \frac{[(M-m+k)(1-\rho_m)+1]\mu_1}{(M-m+k+1)^2}, \tag{15}$$

*2) Statistics of the Optical Channel:* The FSO fading involves three contributions which, are the turbulence-induced fading ($I_a$), the atmospheric path loss ($I_l$) and the pointing errors ($I_p$). The $m$th channel gain $I_m$ can be written as follows:

$$I_m = I_a I_l I_p, \tag{16}$$

Table I summarizes the parameters of the optical part.

Using the Beers-Lambert law, the path loss can be expressed as follows [20, eq. (12)]:

$$I_l = \exp(-\sigma L), \tag{17}$$

The pointing error $I_p$ made by Jitter can be given as [5, eq. (9)]:

$$I_p = A_0 \exp\left(-\frac{2R^2}{\omega_{\text{Leq}}^2}\right), \tag{18}$$

The $\mathcal{M}$-distribution is a generalized model of fading model and it reflects a wide range of turbulences. This fading is eventually based on a physical model involving three components: the first one is the line-of-sight (LOS) component ($U_L$), the second term is quasi-forward scattered by the eddies on the propagation axis coupled to the LOS component ($U_S^C$) and a third component ($U_S^G$) is caused by the energy scattered to the receiver by off-axis eddies. The special cases of the $\mathcal{M}$-distribution are given by [13, Table I].

Since the atmospheric turbulence fading $I_a$ follows the $\mathcal{M}$-distribution, its PDF can be expressed as follows [13, eq. (6)]:

$$f_{I_a}(I_a) = A\sum_{n=1}^{\beta}a_n I_a^{\frac{a+n}{2}-1}K_{\alpha-n}\left(2\sqrt{\frac{\alpha\beta I_a}{g\beta + \Omega'}}\right), \tag{19}$$

where $K_\nu(\cdot)$ is the modified Bessel function of the second kind with order $\nu$, $g = \mathbb{E}[|U_S^C|^2] = 2b_0(1-\rho)$, $2b_0 = \mathbb{E}[|U_S^C|^2 + |U_S^G|^2]$ is the average powers of the LOS, $\Omega = \mathbb{E}[|U_S^L|]$ is the total scatter components, $\alpha$ is a positive parameter related to the effective number of large-scale cells of the scattering process, $\beta$ is a natural number and it stands for the amount of fading parameter, $\Omega' = \Omega + 2\rho b_0 + 2\sqrt{2\rho b_0 \Omega} \cos(\phi_A - \phi_B)$ represents the average power coming from the coherent component, the parameter $\rho$ ($0 \leq \rho \leq 1$) is the amount of scattering power coupled to the LOS component while the parameters $\phi_A$ and $\phi_B$ are the deterministic phases of the LOS and the coupled-to-LOS component. In addition, the parameter $A$ and $a_n$ are defined as:

$$A = \frac{2\alpha^{\frac{\alpha}{2}}}{g^{1+\frac{\alpha}{2}}\Gamma(\alpha)} \left(\frac{g\beta}{g\beta + \Omega'}\right)^{\beta + \frac{\alpha}{2}}, \tag{20}$$

$$a_n = \binom{\beta-1}{n-1} \frac{(g\beta+\Omega')^{1-\frac{n}{2}}}{(n-1)!} \left(\frac{\Omega'}{g}\right)^{n-1} \left(\frac{\alpha}{\beta}\right)^{\frac{n}{2}}, \tag{21}$$

where $\Gamma(\cdot)$ is the incomplete upper gamma function. After unifying the three FSO components, the PDF of the instantaneous SNR $\gamma_{2(m)}$ can be written as [14, eq. (9)]:

$$f_{\gamma_{2(m)}}(\gamma) = \frac{\xi^2 A}{2^r \gamma} \sum_{n=1}^{\beta} b_n G_{1,3}^{3,0}\left(B\left(\frac{\gamma}{\mu_r}\right)^{\frac{1}{r}} \ \middle| \ \begin{matrix}\tau_1\\\tau_2\end{matrix}\right), \tag{22}$$

where $G_{p,q}^{m,n}(\cdot)$ is the Meijer G-function, $\tau_1 = \xi^2 + 1$, $\tau_2 = [\xi^2, \alpha, n]$ and the terms $b_n$ and B are given by:

$$b_n = a_n \left(\frac{g\beta+\Omega'}{\alpha\beta}\right)^{\frac{\alpha+n}{2}}, \tag{23}$$

After some mathematical manipulation, the CDF is given by [14, eq. (11)]:

$$F_{\gamma_{2(m)}}(\gamma) = D \sum_{n=1}^{\beta} c_n G_{r+1,3r+1}^{3r,1}\left(E\frac{\gamma}{\mu_r} \ \middle| \ \begin{matrix}\tau_3\\\tau_4\end{matrix}\right), \tag{24}$$

where $\tau_3 = [1, \ \Delta(r : \xi^2 + 1)]$, $\tau_4 = [\Delta(r : \xi^2), \Delta(r : \alpha), \Delta(r : n), 0]$, $c_n = b_n r^{\alpha+n-1}$, $E = B^r/r^{2r}$, $D = \xi^2 A/[2^r(2\pi)^{r-1}]$ and $\Delta(j; x) \triangleq \frac{x}{j}, \ldots, \frac{x+j-1}{j}$.

Finally, the $k$th moment of the instantenous SNR $\gamma_{2(m)}$ is given by [14, eq. (20)]:

$$\mathbb{E}\left[\gamma_{2(m)}^k\right] = \frac{r_z \xi^2 A \Gamma(rk+\alpha)}{2^r(kr+\xi^2)B^{kr}} \sum_{n=1}^{\beta} b_n \Gamma(rk+n)\mu_r^k, \tag{25}$$

## III. END-TO-END OUTAGE PROBABILITY ANALYSIS

The outage probability is defined as the probability that the end-to-end SNDR falls below an outage threshold $\gamma_{\text{th}}$. It can be written as:

$$P_{\text{out}}(\gamma_{\text{th}}) \triangleq \Pr[\gamma \leq \gamma_{\text{th}}], \tag{26}$$

where $\gamma$ is the effective end-to-end SNDR.

### A. Amplify-and-Forward Relaying

After substituting the expression of the overall SNDR (10) in Eq. (25) and using the following identities [54, eqs. (8.4.3.2) and (2.24.1)], the OP can be derived as follows:

$$P_{\text{out}}(\gamma_{\text{th}}) = 1 - \frac{m\xi^2 A}{2^r(2\pi)^{r-1}} \binom{M}{m} \sum_{k=0}^{m-1} \sum_{n=1}^{\beta} \binom{m-1}{k} e^{-\zeta_1 \frac{\gamma_{\text{th}}}{\mu_1}}$$
$$\times \frac{(-1)^k b_n r^{\alpha+n-1}}{M-m+k+1} G_{r,3r+1}^{3r+1,0}\left(\zeta_2 \frac{\gamma_{\text{th}}}{\mu_1 \mu_r} \ \middle| \ \begin{matrix}\tau_5\\\tau_4\end{matrix}\right), \tag{27}$$

where $\tau_5 = \Delta(r : \xi^2 + 1)$ and $\zeta_1$, $\zeta_2$ are respectively defined as:

$$\zeta_1 = \frac{(M-m+k+1)(1+\kappa_2^2)}{[(M-m+k)(1-\rho_m)+1](1-\delta\gamma_{\text{th}})}, \tag{28}$$

$$\zeta_2 = \frac{(M-m+k+1)C}{[(M-m+k)(1-\rho_m)+1](1-\delta\gamma_{\text{th}})}\left(\frac{B}{r^2}\right)^r, \tag{29}$$

Note that a necessary condition states that the OP is given by Eq. (27) only if $1 - \delta\gamma_{\text{th}} > 0$, otherwise it is equal to a unity.

To derive the asymptotical high SNR, we refer to the expansion of the Meijer-G function as follows [55, eq. (07.34.06.0001.01)]:

$$G_{r+1,3r+1}^{3r+1,0}\left(\zeta_2 \frac{\gamma_{\text{th}}}{\mu_1 \mu_r} \ \middle| \ \begin{matrix}\tau_5\\\tau_4\end{matrix}\right)$$
$$\underset{\mu_r \gg 1}{\cong} \sum_{v=1}^{3r+1} \frac{\prod_{j=1, j\neq v}^{3r+1} \Gamma(\tau_{4,j} - \tau_{4,v})}{\prod_{j=1}^{r} \Gamma(\tau_{5,j} - \tau_{4,v})} \times \left(\frac{\zeta_2 \gamma_{\text{th}}}{\mu_1 \mu_r}\right)^{\tau_{4,v}}, \tag{30}$$

### B. Decode-and-Forward Relaying

Without loss of generality, we assume that all relays are able to decode the received signals from the source (S). For non-ideal hardware, the OP for DF relaying can be expressed as follows:

$$P_{\text{out}}(\gamma_{\text{th}}) = 1 - \prod_{i=1}^{2} \left(1 - F_{\gamma_{i(m)}}\left(\frac{\gamma_{\text{th}}}{1-\kappa_i^2\gamma_{\text{th}}}\right)\right), \tag{31}$$

Note that the CDFs $F_{\gamma_{i(m)}}$, $i = 1, 2$ are defined only if $\gamma_{\text{th}} < \frac{1}{\kappa_i^2}$, i.e. $\gamma_{\text{th}} < \frac{1}{\max(\kappa_1^2, \kappa_2^2)} = \frac{1}{\delta}$. Hence, the OP is given by Eq. (30) if $\gamma_{\text{th}} < \frac{1}{\delta}$, otherwise it is equal to a unity.

In case of ideal hardware, the expression of the OP is reduced to:

$$P_{\text{out}}(\gamma_{\text{th}}) = 1 - \prod_{i=1}^{2} \left(1 - F_{\gamma_{i(m)}}(\gamma_{\text{th}})\right), \tag{32}$$

Similar to (27) and (30), the asymptotical high SNR can be derived by using the expansion of the Meijer-G function

as follows:

$$F_{\gamma_{2(m)}}\left(\frac{\gamma_{\text{th}}}{1-\kappa_2^2\gamma_{\text{th}}}\right)\underset{\mu_r\gg1}{\cong} D\sum_{n=1}^{\beta}\sum_{v=1}^{3r}\frac{\prod_{j=1,j\neq v}^{3r}\Gamma(\tau_{4,j}-\tau_{4,v})}{\prod_{j=2}^{r+1}\Gamma(\tau_{5,j}-\tau_{4,v})}$$
$$\times\frac{\Gamma(\tau_{4,v})}{\Gamma(1+\tau_{4,v})}\left(\frac{E\gamma_{\text{th}}}{(1-\kappa_2^2\gamma_{\text{th}})\mu_r}\right)^{\tau_{4,v}},\tag{33}$$

## IV. SYMBOL ERROR PROBABILITY ANALYSIS

For the most coherent linear modulation, the SEP is provided as follows:

$$\overline{P_e}=\mathbb{E}\left[\mathcal{Q}(\sqrt{c\gamma})\right],\tag{34}$$

where $\mathcal{Q}(\cdot)$ is the Gaussian-$\mathcal{Q}$ function and $c$ is a parameter related to the format of the modulation, e.g, $c=2$ stands for BPSK modulation. After applying an integration by parts on Eq. (32), SEP can be expressed as follows:

$$\overline{P_e}=\sqrt{\frac{c}{8\pi}}\int_0^\infty\frac{e^{-\frac{c}{2}\gamma}}{\sqrt{\gamma}}F_\gamma(\gamma)d\gamma,\tag{35}$$

### A. Amplify-and-Forward Relaying

Because of the terms related to the hardware impairments, the derivation of the average SEP is not tractable. In that way, a numerical integration is required to plot the variations of the average SEP. Since the end-to-end OP is constrained by the necessary condition, the upper bound of the integral defined in Eq. (33) is equal to $1/\delta$.

For the ideal case ($\kappa_1=\kappa_2=0$) and after using the following identities [56, eq. (3.381.4)] and [54, eq. (2.24.3.1)], the average SEP can be easily derived as follows:

$$\overline{P_e}=\frac{1}{2}-\sqrt{\frac{c}{8\pi}}\frac{m\zeta^2A}{2^r(2\pi)^{r-1}}\binom{M}{m}\sum_{k=0}^{m-1}\sum_{n=1}^{\beta}\binom{m-1}{k}$$
$$\times\frac{(-1)^kb_nr^{\alpha+n-1}}{M-m+k+1}\sqrt{\frac{2\mu_1}{c\mu_1+2\zeta_1}}$$
$$\times G_{r+1,3r+1}^{3r+1,1}\left(\frac{2\zeta_2}{(c\mu_1+2\zeta_1)\mu_r}\;\middle|\;\begin{matrix}\tau_6\\\tau_4\end{matrix}\right),\tag{36}$$

where $\tau_6=[0.5,\;\tau_5]$. A high SNR asymptote can be derived by using the expansion of the Meijer-G function as follows [55, eq. (07.34.06.0001.01)]:

$$G_{r+1,3r+1}^{3r+1,1}\left(\frac{2\zeta_2}{(c\mu_1+2\zeta_1)\mu_r}\;\middle|\;\begin{matrix}\tau_6\\\tau_4\end{matrix}\right)\underset{\mu_r\gg1}{\cong}$$
$$\times\sum_{v=1}^{3r+1}\frac{\prod_{j=1,j\neq v}^{3r+1}\Gamma(\tau_{4,j}-\tau_{4,v})\Gamma(0.5+\tau_{4,v})}{\prod_{j=2}^{r+1}\Gamma(\tau_{6,j}-\tau_{4,v})}$$
$$\times\left(\frac{2\zeta_2}{(c\mu_1+2\zeta_1)\mu_r}\right)^{\tau_{4,v}},\tag{37}$$

### B. Decode-and-Forward Relaying

Similar to the case of the AF relaying, the derivation of the average SEP is complex due to the presence of the terms related to the hardware impairments. Hence, a numerical integration is required. The impact of the hardware impairments on the system performance is the creation of an irreducible floor that saturates the average SEP. Hence, the diversity gain $G_d$ is equal to zero. For the ideal case, since the CDF of the instantaneous SNR consists of complex functions such the Meijer-G function, they do not provide insight into the system performance. Consequently, it is more meaningful to derive the average SEP at high SNR range as follows:

$$\overline{P_e}\approx(G_c\overline{\gamma})^{-G_d},\tag{38}$$

where $G_d$ and $G_c$ are the diversity and the coding gains. To achieve this step, we use the technique proposed by [57]–[60] to approximate the PDF of the overall SNR as follows:

$$f_\gamma(\gamma)=a\gamma^b+o(\gamma),\tag{39}$$

From the above approximation, the asymptotical expression of the average SEP can be formulated as follows:

$$\overline{P_e}\approx\frac{\prod_{i=1}^{b+1}(2i-1)}{2(b+1)!c^{b+1}}\frac{\partial^bf_\gamma}{\partial\gamma^b}(0)=\frac{2^ba\Gamma(b+3/2)}{\sqrt{\pi}(b+1)}(c\overline{\gamma})^{-(b+1)},\tag{40}$$

where $a$ is a constant and $b$ must be a natural numer for the first equation in (37) and not necessarily an integer for the second equation. As a result, it is required first to derive the approximate PDF to find the diversity order $G_d=b+1$ and the coding gain $G_c$. Given that the CDF of the overall SNR for the ideal case is given by Eq. (31), it can be approximated at high SNR region as follows:

$$F_\gamma(\gamma)\approx F_{\gamma_{1(m)}}(\gamma)+F_{\gamma_{2(m)}}(\gamma),\tag{41}$$

After deriving (39), the approximate PDF of the end-to-end SNR is given by:

$$f_\gamma(\gamma)\approx f_{\gamma_{1(m)}}(\gamma)+f_{\gamma_{2(m)}}(\gamma),\tag{42}$$

Since $\gamma_{1(m)}$ is exponentially distributed under the assumption of PRS with outdated CSI, $b$ is equal to zero. On the other side, the high SNR approximation of $f_{\gamma_{2(m)}}$ can be derived by using the expansion of the Meijer-G function as follows:

$$f_{\gamma_{2(m)}}(\gamma)\approx\frac{\zeta^2A}{2^r\gamma}\sum_{n=1}^{\beta}\sum_{v=1}^{3}\frac{\prod_{j=1,j\neq v}^{3}\Gamma(\tau_{2,j}-\tau_{2,v})}{\Gamma(\xi^2+1-\tau_{2,v})}$$
$$\times B^{\tau_{2,v}}\left(\frac{\gamma}{\mu_r}\right)^{\frac{\tau_{2,v}}{r}},\tag{43}$$

Therefore, the PDF of $\gamma_{2(m)}$ can be reformulated as follows:

$$f_{\gamma_{2(m)}}(\gamma)\approx D\gamma^{\min\left(\frac{\xi^2}{r},\frac{\alpha}{r},\frac{\beta}{r}\right)},\tag{44}$$

where D is a constant parameter. After combining the PDF approximations of $\gamma_{1(m)}$ and $\gamma_{2(m)}$, the PDF of the overall SNR can be derived as follows:

$$f_\gamma(\gamma) \approx a\gamma^{\min\left(1,\ \min\left(\frac{\xi^2}{r},\ \frac{a}{r},\ \frac{\beta}{r}\right)\right)}, \qquad (45)$$

Finally, the diversity gain $G_d$ can be given by:

$$G_d = \min\left(1,\ \min\left(\frac{\xi^2}{r},\ \frac{a}{r},\ \frac{\beta}{r}\right)\right), \qquad (46)$$

While the array gain $G_c$ can be derived as follows:

$$G_c = c\left(\frac{2^b a\Gamma(a+3/2)}{\sqrt{\pi}(b+1)}\right)^{-\frac{1}{b+1}}, \qquad (47)$$

## V. ERGODIC CAPACITY ANALYSIS

In this section, we will provide the analysis of the ergodic capacity, expressed in bps/Hz. It is defined as the maximum error-free data rate transferred by the system channel.

### A. Capacity of AF Relaying

The ergodic capacity of AF relaying protocol with ideal hardware has been properly investigated in the literature [61]–[63]. Considering the case of hardware impairments, the channel capacity can be written as follows:

$$\overline{C} \triangleq \mathbb{E}\left[\log_2(1+\varpi\,\gamma)\right], \qquad (48)$$

where $\varpi = 1$ indicates the heterodyne detection and $\varpi = \frac{e}{2\pi}$ for IM/DD. The ergodic capacity can be derived by calculating the PDF of the SNDR. However, an exact analytical formulation of Eq. (46) is very complex due to the presence of the terms related to the hardware impairments. To calculate the ergodic capacity, we should refer to the numerical integration method.

To quantify the EC, there upper bound which, is given by the following theorem.

*Theorem 1:* For Asymmetric (Rayleigh/Málaga) fading channels, the ergodic capacity $\overline{C}$ with AF relaying and non-ideal hardware has an upper bound defined by:

$$\overline{C} \leq \log_2\left(1 + \frac{\varpi\,\mathcal{I}}{\delta\mathcal{I}+1}\right), \qquad (49)$$

where $\mathcal{I}$ is defined as:

$$\mathcal{I} \triangleq \mathbb{E}\left[\frac{\gamma_{1(m)}\gamma_{2(m)}}{(1+\kappa_2^2)\gamma_{2(m)}+C}\right], \qquad (50)$$

After some mathematical manipulations, $\mathcal{I}$ can be derived as follows:

$$\mathcal{I} = \frac{\xi^2 A\mathbb{E}\left[\gamma_{1(m)}\right]}{2^r(2\pi)^{r-1}(1+\kappa_2^2)}\sum_{n=1}^{\beta} b_n r^{a+r-1}$$
$$\times G_{r+1,3r+1}^{3r+1,1}\left(\left(\frac{B}{r^2}\right)^r \frac{C}{(1+\kappa_2^2)\mu_r}\ \Big|\ \begin{matrix}\tau_5,\ 0\\ \tau_4\end{matrix}\right), \qquad (51)$$

*Proof:* The proof is given in Appendix A.  ∎

Although deriving a closed-form of the ergodic capacity is very complex, we can find an approximate simpler form

by applying the approximation given by [48, eq. (27)], [49, eq. (35)]:

$$\mathbb{E}\left[\log_2\left(1+\frac{\psi}{\varphi}\right)\right] \approx \log_2\left(1+\frac{\mathbb{E}\left[\psi\right]}{\mathbb{E}\left[\varphi\right]}\right), \qquad (52)$$

Now, let's consider a practical and realistic model of hardware impairment to test the resiliency of the proposed relaying system. We assume an ideal source and impaired relays that suffer from non-linear high power amplifier (HPA). We suggest the two HPA models called Soft Envelope Limiter (SEL) and Traveling Wave Tube Amplifier (TWTA) proposed by [48] and [64]. In this case, the overall SNDR can be written as follows [64, eq. (24)]:

$$\gamma = \frac{\gamma_{1(m)}\gamma_{2(m)}}{\kappa\gamma_{2(m)}+C}, \qquad (53)$$

where $C = \mathbb{E}\left[\gamma_{1(m)}\right] + \kappa$ and $\kappa$ is given by [64, eq. (25)]:

$$\kappa = 1 + \frac{\sigma_\tau^2}{\varepsilon^2 G^2\sigma_0^2}, \qquad (54)$$

The parameters $\varepsilon$ and $\sigma_\tau^2$ are derived for SEL and TWTA in [64, eqs. (18) and (19)], respectively. We also define the Input Back-Off (IBO) relative to the amplifier as follows:

$$\text{IBO} = \frac{A_{\text{sat}}^2}{\sigma_\tau^2}, \qquad (55)$$

where $A_{\text{sat}}$ is the saturation level of the relay's amplifier. Note that the ideal hardware case can be achieved only if $\kappa$ converges to one.

In order to derive the average ergodic capacity, we should first find the expression of the CDF of overall SNR. Similar to the derivation steps used to find the OP (27), the CDF of the end-to-end SNDR can be written as follows:

$$F_\gamma(\gamma_{\text{th}}) = 1 - \frac{m\xi^2 A}{2^r(2\pi)^{r-1}}\binom{M}{m}\sum_{k=0}^{m-1}\sum_{n=1}^{\beta}\binom{m-1}{k}e^{-\zeta_3\frac{\gamma_{\text{th}}}{\mu_1}}$$
$$\times \frac{(-1)^k b_n r^{a+n-1}}{M-m+k+1}G_{r,3r+1}^{3r+1,0}\left(\zeta_4\frac{\gamma_{\text{th}}}{\mu_1\mu_r}\ \Big|\ \begin{matrix}\tau_5\\ \tau_4\end{matrix}\right), \qquad (56)$$

where $\zeta_3$ and $\zeta_4$ are given by:

$$\zeta_3 = \frac{(M-m+k+1)\kappa}{[(M-m+k)(1-\rho_m)+1]}, \qquad (57)$$

$$\zeta_4 = \frac{(M-m+k+1)C}{[(M-m+k)(1-\rho_m)+1]}\left(\frac{B}{r^2}\right)^r, \qquad (58)$$

After applying the inetgration by parts on (46), the ergodic capacity can be reformulated as follows:

$$\overline{C} = \frac{\varpi}{\ln(2)}\int_0^\infty (1+\varpi\gamma)^{-1}\overline{F}_\gamma(\gamma)\,d\gamma, \qquad (59)$$

where $\overline{F}_\gamma(\cdot)$ is the complementary CDF (CCDF) of $\gamma$. After replacing the CCDF of (53) in (56), the ergodic capacity can be derived in closed-form as follows:

$$
\begin{aligned}
\overline{C} = {} & \frac{m\xi^2 A \varpi \mu_1}{2^r (2\pi)^{r-1} \ln(2)} \binom{M}{m} \sum_{k=0}^{m-1} \sum_{n=1}^{\beta} \binom{m-1}{k} \\
& \times \frac{(-1)^k b_n r^{\alpha+n-1}}{(M-m+k+1)\zeta_3} H_{1,0:1,1;r,3r+1}^{0,1:1,1:3r+1,0} \\
& \times \begin{pmatrix} (0;1,1) & \Big| & (0,1) & \Big| & (\tau_5,[1]_r) \\ - & \Big| & (0,1) & \Big| & (\tau_6,[1]_{3r+1}) \end{pmatrix} \frac{\varpi \mu_1}{\zeta_3}, \frac{\zeta_4}{\zeta_3 \mu_r} \Big),
\end{aligned}
\tag{60}
$$

where $H_{p_1,q_1:p_2,q_2:p_3,q_3}^{m_1,n_1:m_2,n_2:m_3,n_3}(-|(\cdot,\cdot))$ is the bivariate Fox-H function and $[x]_j$ is the vector containing $j$ elements equal to $x$.

An efficient Matlab implementation of the bivariate Fox-H function is provided in [65].

*Proof:* The proof is given in Appendix B. ∎

To derive the High SNR approximation of the ergodic capacity, we should expand the Meijer-G function in the expression (53) and then replace it (CCDF) in Eq. (56). After using the identities [54, eqs. (8.4.2.5) and (2.24.3.1)] and [55, eq. (07.34.06.0001.01)], the ergodic capacity can be expressed at high SNR region as follows:

$$
\begin{aligned}
\overline{C} \underset{\mu_r \gg 1}{\cong} {} & \frac{m\xi^2 A}{\ln(2) 2^r (2\pi)^{r-1}} \binom{M}{m} \sum_{k=0}^{m-1} \sum_{n=1}^{\beta} \sum_{\nu=1}^{3r+1} \binom{m-1}{k} \\
& \times \frac{(-1)^k b_n r^{\alpha+n-1}}{M-m+k+1} \left(\frac{\mu_1}{\zeta_3}\right)^{\tau_{4,\nu}+1} \left(\frac{\zeta_4}{\mu_1 \mu_r}\right)^{\tau_{4,\nu}} \\
& \times \frac{\prod_{j=1, j\neq \nu}^{3r+1} \Gamma(\tau_{4,j} - \tau_{4,\nu}) \Gamma(1+\tau_{4,\nu})}{\prod_{j=1}^{r} \Gamma(\tau_{5,j} - \tau_{4,\nu})},
\end{aligned}
\tag{61}
$$

### B. Capacity of DF Relaying

The analysis of the ergodic capacity for the DF relaying is more complicated than the AF relaying protocol. Unlike the AF relaying in term of complexity of processing, the relay must decode and re-encode the information signal which, rises the level of difficulties. If the relay fails to decode the signal, the transmitted signal will be useless. In case of successfull decoding, the relay re-encodes the information and allocates the resources of power and symbols required for the transmission. To avoid further calculus constraints, we assume that all relays are able to decode the signal so that the complexity of processing will be reduced. In addition, we also assume that all relays are assigned equal times of resources allocation and prescheduling.

According to [66, eq. (45)] and [61, eq. (11a)] and the expression of the end-to-end SNDR in Eq. (10), the upper bound of the ergodic capacity with accounting of hardware impairments [49, eq. (38)] can be written as follows:

$$
\overline{C} \leq \min_{i=1,2} \mathbb{E}\left[\log_2\left(1 + \frac{c\gamma_{i(m)}}{\kappa_i^2 \gamma_{i(m)} + 1}\right)\right],
\tag{62}
$$

Eq. (51) shows clearly the effects of the hardware impairments in limiting the system performance contrary to the case of the ideal hardware where the system capacity grows infinitely.

### C. Asymptotic Analysis

For high SNR regime, it is trivial that the end-to-end effective SNDR $\gamma$ for AF and DF relaying converges to a ceiling $\gamma^*$ defined as:

$$
\gamma^* \triangleq \begin{cases} \dfrac{1}{\kappa_1^2 + \kappa_2^2 + \kappa_1^2\kappa_2^2} & \text{AF Relaying Protocol,} \\[2ex] \dfrac{1}{\max(\kappa_1^2, \kappa_2^2)} & \text{DF Relaying Protocol,} \end{cases}
\tag{63}
$$

We observe that $\gamma^*$ is inversely proportional to the hardware impairments $\kappa_1^2$ and $\kappa_2^2$. This confirms that the hardware impairments deeply affect the system performance and so it must be considered in the system modeling. We also observe that the ceiling for DF relaying is half of the ceiling for AF relaying protocol. This implies that the DF relaying is more resilient to the hardware impairments than the AF protocol. Regarding the ergodic capacity in a high SNR regime, the hardware impairments saturate the channel capacity. This fact is shown by the following corollary.

*Corollary 1:* Suppose that $\mu_1$, $\mu_r$ largely increase and the electrical and optical channels are mutually independant with strictly non-negative fading, the ergodic capacity converges to a capacity ceiling defined by $\overline{C^*} = \log_2(1 + \varpi \gamma^*)$.

*Proof:* Since the SNDR converges to $\gamma^*$, the dominated convergence theorem consequently allows to move the limit inside the logarithm function as shown below:

$$
\begin{aligned}
\lim_{\mu_1,\mu_r \to \infty} \log_2(1 + \varpi\, \mathbb{E}[\gamma]) &= \log_2(1 + \varpi \lim_{\mu_1,\mu_r \to \infty} \mathbb{E}[\gamma]) \\
&= \log_2(1 + \varpi \gamma^*),
\end{aligned}
\tag{64}
$$

Hence, the ergodic capacity with AF relaying protocol under harware impairments satisfies:

$$
\lim_{\mu_1,\mu_r \to \infty} \overline{C} = \log_2\left(1 + \frac{\varpi}{\kappa_1^2 + \kappa_2^2 + \kappa_1^2\kappa_2^2}\right),
\tag{65}
$$

For DF relaying and non-ideal harware impairments, the ergodic capacity is upper bounded by:

$$
\lim_{\mu_1,\mu_r \to \infty} \overline{C} \leq \log_2\left(1 + \frac{\varpi}{\max(\kappa_1^2, \kappa_2^2)}\right),
\tag{66}
$$

If we assume that only the relays are susceptible to the non-linear HPA impairment, the ergodic capacity is saturated by a ceiling that depends only on the amplifier's parameters as follows [48, eq. (37)]:

$$
\overline{C_c} = \log_2\left(1 + \frac{\varpi \varepsilon^2}{\iota - \varepsilon^2}\right),
\tag{67}
$$

where $\iota$ is called the clipping factor. Further details about the non-linear HPA modeling are found in [32] and [64].

## VI. NUMERICAL RESULTS

This section provides numerical results obtained by using the mathematical formulations of the previous section. The electrical channel is subject to the correlated Rayleigh fading which, can be generated using the algorithm in [67]. The turbulence-induced fading is modeled by $\mathcal{M}$-distribution, which, can be generated by using the formula [13, eq. (2)], $I = XY$, where $X \frown \mathcal{G}(\alpha, 1)$ and $Y \frown \mathcal{SR}(g, \beta, \rho, \Omega', \Delta\Phi)$ are mutually independent random variables. In addition, the pointing error is simulated by generating the radial displacement $R$ following the Rayleigh distribution and then applying Eq. (18). Since the path loss is deterministic, it can be generated using relation (17). Unless otherwise stated, Table II presents the main simulation parameters.

| Parameter | Value |
|---|---|
| $C_n^2$ | $2.8 \ 10^{-14}$ m$^{-2/3}$ |
| $L$ | 1 km |
| $\lambda$ | 1550 nm |
| $\gamma_{th}$ | 7 dB |
| $F_0$ | -10 m |
| $a$ | 5 cm |
| $w_0$ | 5 mm |
| $b_0$ | 0.596 |
| $\Omega$ | 1.32 |
| $\alpha$ | 4.2 |
| $\beta$ | 5 |
| $\rho$ | 0.6 |
| $\rho_m$ | 0.7 |
| M, m | 3 |
| $\kappa_1$, $\kappa_2$ | 0.3 |
| Modulation | CBPSK |

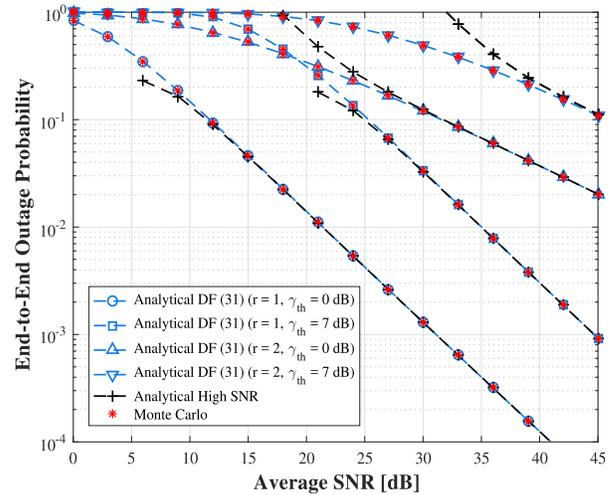

Fig. 3. Outage Probability for IM/DD and heterodyne detections using different $\gamma_{th}$.

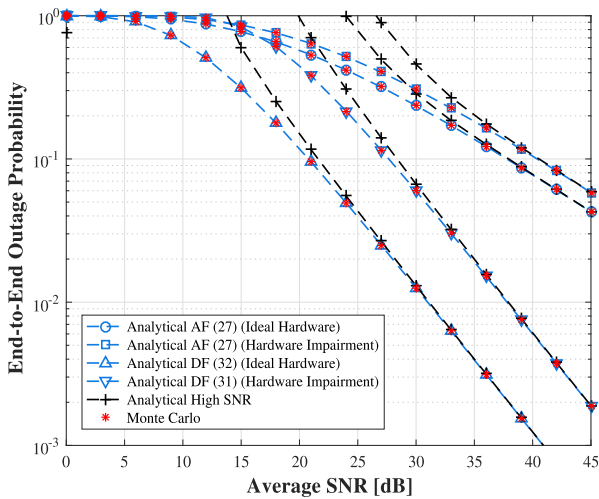

Fig. 2. Outage Probability for ideal and non-ideal hardware under IM/DD detection.

Fig. 2 shows the OP dependence on the average SNR for AF and DF relaying protocols. We note that for low SNR, the performance of ideal and non-ideal hardware shows a slight deviation from each other. In this case, the hardware impairments has a small impact on the outage performance and so the assumption of neglecting the effect of the impairments can be valid. As the SNR increases toward 30 dB, the performance deteriorates and so the approximation of neglecting the hardware impairments is no longer valid. Although the impairments factor has an inavoidable effect on the system performance, the DF relaying protocol appears to be more resilient than the AF protocol. In fact, we observe that even though a DF relaying system operates under hardware impairments, it outperforms an AF relaying system with ideal hardware. This result as expected since the distortion noise of the first RF channels is carried on the second optical channels for AF protocol.

Fig. 3 shows the OP performance as a function of the average SNR for AF relaying protocol. The curve variations show that there is a small performance loss caused by the hardware impairments for the low threshold $\gamma_{th} = 2$ dB. Whereas, there are substantial losses when the outage threshold increases to 5 dB. Regarding the detection method, the graph is absolutely in agreement with previous work. As expected, the relaying system works better with heterodyne detection than using IM/DD method.

The dependence of the OP for DF relaying protocol on the average SNR is given by Fig. 4. As expected, the outage performance is better under the moderate turbulence condition and suddenly deteriorate as the turbulence becomes strong and severe. This result is clearly observed, especially for the case of full correlation of CSIs ($\rho = 1$). It turned out that the system substantially depends on the state of the optical channels. As the correlation $\rho$ between the CSI used for relay selection and the CSI used for transmission increases, i.e., the two CSIs become more and more correlated, the selection of the best relay is certainly achieved ($m = M$). In this case, the system works under the perfect condition specially under moderate turbulence condition. As the time correlation decreases, the selection of the best relay is no longer achieved and so the system certainly operates with a worse relay.

Fig. 5 shows the OPs of AF and DF relaying protocols as a function of the threshold $\gamma_{th}$ (dB) for ideal and non-ideal hardware ($\kappa_1 = \kappa_2 = 0.3$). For small outage threshold, the OPs are slightly deteriorated by the hardware impairments. For high outage threshold, the system with ideal harware smoothly converges toward 1, while the non-ideal system is subject to a rapid convergence to the SNDR ceiling. As we concluded before, DF relaying protocol is more robust to hardware impairments and the relative SNDR ceiling is higher than of AF relaying scheme. For practical use when dealing with non-ideal hardware, DF relaying sufficiently proves its superiority over AF relaying scheme to acquire the hardware imperfections caused by low quality of the materials.

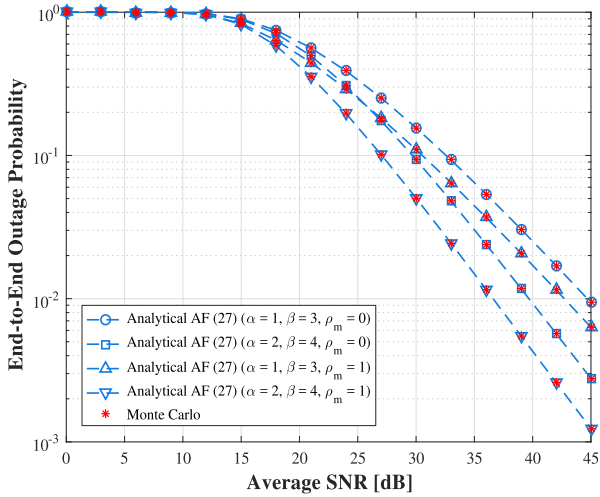

Fig. 4. Outage probability for various correlation and turbulences.

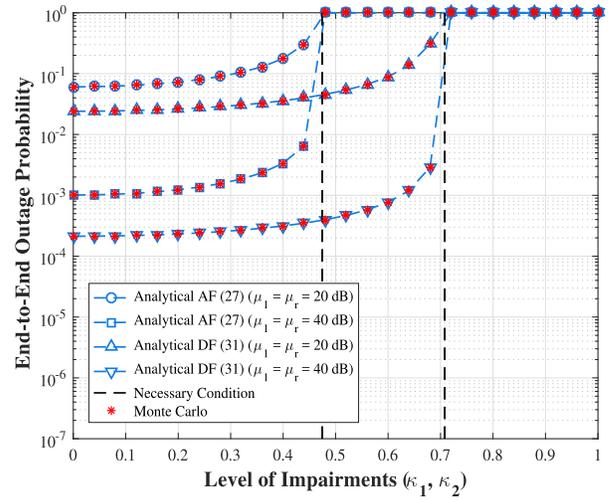

Fig. 6. Outage probability for various levels of hardware impairments.

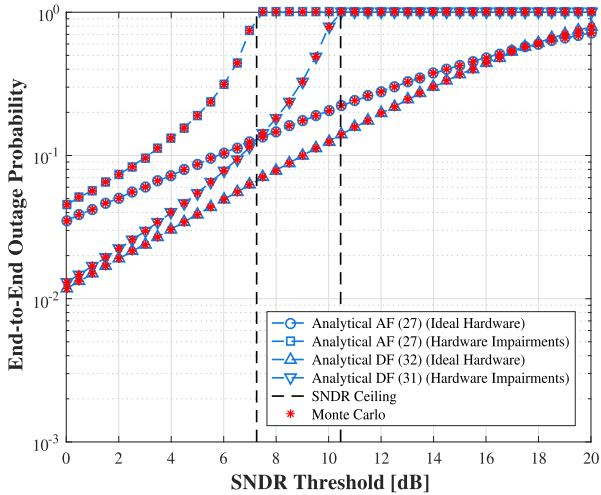

Fig. 5. Outage probability versus the SNDR threshold for ideal and non-ideal hardware.

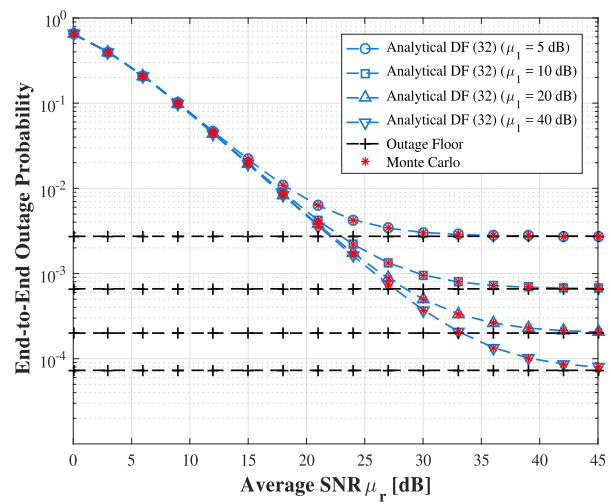

Fig. 7. Outage probability versus the average optical SNR for different $\mu_1$.

The variations of the OP for AF and DF protocols with respect to the level of the impairments for two different average SNRs $\mu_1 = \mu_2 = \{20, 40\}$dB are shown in Fig. 6. Considering the case 40 dB and requiring that the OP is under $10^{-2}$, we can identify two operating regimes:

1) AF relaying protocol with $\kappa_1 = \kappa_2 \leq 0.44$.
2) DF relaying protocol with $\kappa_1 = \kappa_2 \leq 0.70$.

The various acceptable levels of impairments prove that DF relaying protocol is more resistant to the hardware impairments and thus, it can operate with low quality of the hardware for practical use. Fig. 6 also shows the necessary condition we mentioned earlier,, is an upper bound on the impairments level that can achieve an OP less than one.

Fig. 7 shoes the impact of the average RF SNR on the outage performance. In fact, the system saturates as the average transmitted power over the first hop is constant. The limitation is mainly observed by the creation of the outage floor that substantially degrades the system performance. Another important metric that is considered is the error performance.

The degradation caused by the hardware impairments is confirmed again by the saturation of the error performance shown by Fig. 8. We observe the creation of an error floor that limits the system performance and this floor becomes more severe as the average SNR increases.

The error performance is also investigated for different weather states in Fig. 9. For clear air, the weather is quiet and the scattering loss is negligible or small. Given that the high frequency signals are greatly disturbed by the fog, clouds and dust particles, the FSO signal depends not only on the rain which, is the major attenuating factor but also on the rate of the rainfall as shown by the figure. In fact, the rain droplets cause a substantial scattering in different directions that mainly attenuate the signal power during the propagation

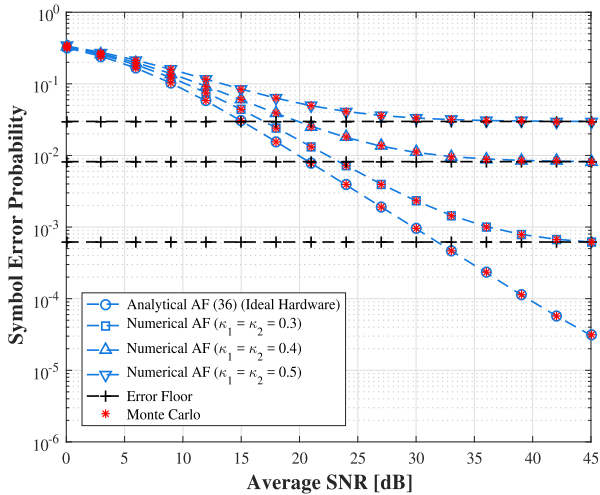

Fig. 8.   Symbol error probability for various levels of hardware impairments.

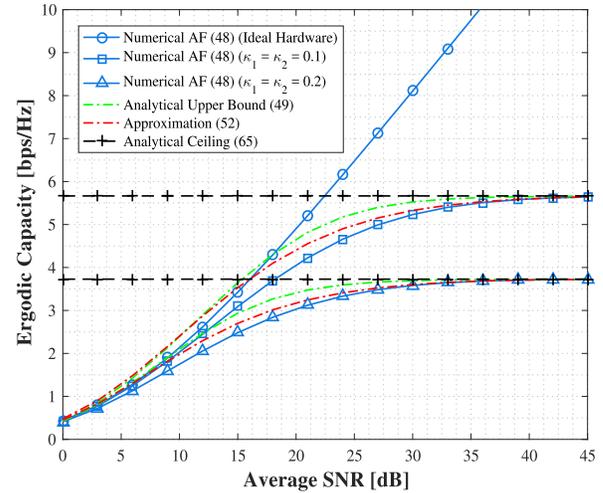

Fig. 10.   Ergodic capacity for different values of the hardware impairments.

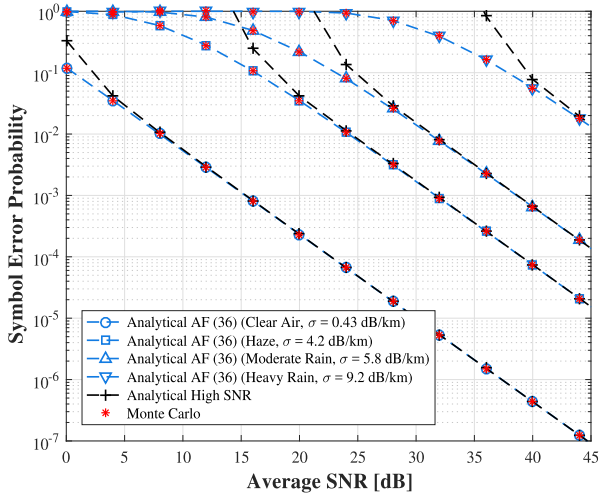

Fig. 9.   Symbol error probabilty probability for various weather attenuation coefficients.

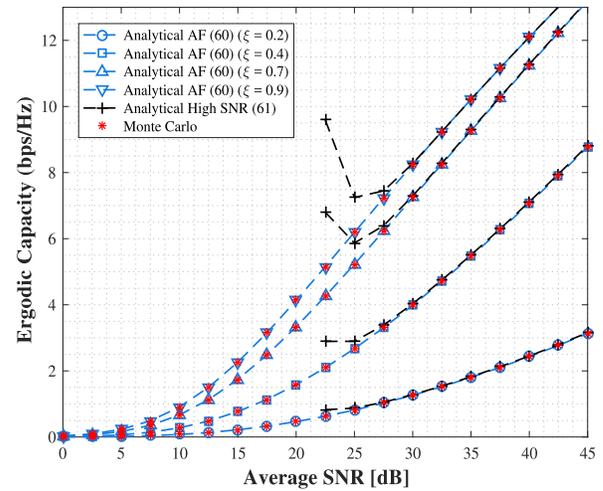

Fig. 11.   Ergodic capacity for different values of the pointing error coefficients.

and this phenomena can be explained in more details according to the Rayleigh model of scattering.

Fig. 10 shows the dependence of the ergodic capacity on the average SNR for ideal and non-ideal harware. We observe that the hardware impairments level is acceptable at low SNR, but it becomes very severe as the SNR increases. Specifically, the ergodic capacity saturates and converges to the capacity ceiling $C^*$ as shown by corollary 1, which, is inversely proportional to $\kappa_1$ and $\kappa_2$. Fig. 10 also presents the capacity upper bound proved by Theorem 1 and the approximation (52). Although the exact, the approximate and the upper bound of the ergodic capacity show a slight deviation from each other at low SNR regime, they are asymptotically exact as the average SNR largely increases.

Fig. 11 provides the variations of the ergodic capacity against the average SNR for different values of the pointing error coefficients. We observe that the system works better as the pointing error coefficient decreases. In fact, as this coefficient $\xi$ decreases, the pointing error effect becomes more severe. For a given average SNR of 30 dB, the system capacity achieves the following rates 1, 3.9, 7 and 8 bps/Hz for the pointing error coefficients equal to 0.2, 0.4, 0.7 and 0.9, respectively. Thereby, the ergodic capacity gets better as the pointing error coefficient becomes higher.

Fig. 12 shows the variations of the ergodic capacity versus the average SNR for different values of IBO. Clearly, we observe that the ergodic capacity saturates by the ceilings that are caused by the hardware impairments as shown by the figure. In addition, these ceilings disappear for an IBO = 30 dB as shown in Fig. 3 but the performances are limited for the case of lower values of IBO. For the following values of IBO equaling to 0, 3, 5 and 7 dB, the system capacity is saturated by the following ceiling values 3, 4.9, 6.6 and 9.8 bps/Hz, respectively. Note that these ceilings are inversely proportional to the values of the IBO. In fact, as the IBO increases, the saturation amplitude of the relay amplifier increases and thus the distortion effect is reduced. However, as the IBO decreases, i.e, the relay amplifier level becomes lower, the non-linear distortion impact becomes more severe and the channel capacity substantially saturates. Note that the capacity ceiling depends only on the hardware impairment parameters like the clipping factor and the scale of the input signal and not on the system parameters such as the number of relays and the channels' parameters, etc.

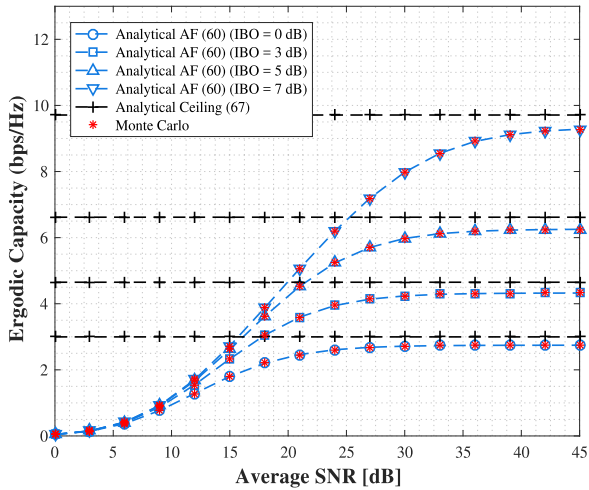

Fig. 12. Ergodic capacity of AF relaying for different values of the relay's amplifier IBO.

## VII. CONCLUSION

In this work, we introduced a general model of impairments to a mixed RF/FSO system for AF and DF relaying protocols and we observed that it has a small impact on the system for low SNRs but this effect becomes more severe as the SNR increases. In addition, we proved that the correlation has a deep impact on the system performance but importantly, the system depends to a large extent on the state of the optical channel in terms of the turbulence intensity, the weather attenuation and the pointing error. Furthermore, we proved that the DF relaying is more efficient than AF protocol and so it is more convenient for practical use. Also, we investigated the impact of the non-linear HPA on the system performance and we concluded that the system works better for higher values of IBO.

## APPENDIX A
## DERIVATION OF THE TERM $\mathcal{I}$

$\log_2\left(1 + \frac{\varpi\,\mathcal{I}}{\delta\mathcal{I}+1}\right)$ is concave of $\mathcal{I}$ for $\mathcal{I} > 0$ since its second derivative is given by:

$$\frac{-(2\delta^2\mathcal{I} + 2\delta(\mathcal{I}+1) + 1)}{\log_e(2)(\delta\mathcal{I}+1)^2(\delta\mathcal{I}+\mathcal{I}+1)^2} < 0, \tag{68}$$

Then, we can apply the Jensen's inequality to get:

$$\overline{C} \leq \log_2\left(1 + \frac{\varpi\,\mathcal{I}}{\delta\mathcal{I}+1}\right), \tag{69}$$

After applying the following identities [54, eqs. (8.4.2.5) and (2.24.1)] and using (15), the term $\mathcal{I}$ is finally derived as given by (51).

## APPENDIX B
## DERIVATION OF THE ERGODIC CAPACITY
## UNDER THE NON-LINEAR HPA (60)

The first step is to refer to the following identities [54, eqs. (8.4.2.5) and (8.4.3.1)] to transform both the fraction and the exponential into Meijer-G function, respectively. After the previous transformation, the integral involves three Meijer-G functions that each one of them should be transformed into the Fox-H function using the following formula:

$$G_{p,q}^{m,n}\left[z^C\,\middle|\,\begin{matrix}a_1,\ldots,a_p\\b_1,\ldots,b_q\end{matrix}\right]$$
$$= \frac{1}{C}\,H_{p,q}^{m,n}\left[z\,\middle|\,\begin{matrix}(a_1,C^{-1}),\ldots,(a_p,C^{-1})\\(b_1,C^{-1}),\ldots,(b_q,C^{-1})\end{matrix}\right], \tag{70}$$

After solving the integral using [68, eq. (2.3)], the ergodic capacity is finally derived in terms of the bivariate Fox-H function as given by (60).

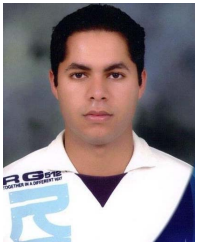

**Elyes Balti** (S'16) was born in Jendouba, Tunisia. He received the B.S. degree in electrical engineering from the Ecole Supérieure des Communications de Tunis (Sup'Com), Tunisia, in 2013. He is currently pursuing the M.Eng. degree in electrical engineering with the University of Idaho, USA. His research interests include wireless communications and networking, including the millimeter wave and optical wireless channel modeling, 5G cellular, and MIMO communication systems.

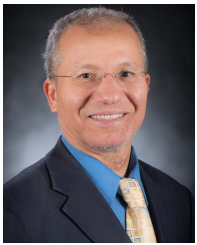

**Mohsen Guizani** (S'85–M'89–SM'99–F'09) received the B.S. (Hons.) and M.S. degrees in electrical engineering, and the M.S. and Ph.D. degrees in computer engineering from Syracuse University, Syracuse, NY, USA, in 1984, 1986, 1987, and 1990, respectively. He served as the Associate Vice President of Graduate Studies, Qatar University, the Chair of the Computer Science Department, Western Michigan University, and the Chair of the Computer Science Department, University of West Florida. He also served in academic positions at the University of Missouri–Kansas City, the University of Colorado Boulder, Syracuse University, and Kuwait University. He is currently a Professor and the ECE Department Chair with the University of Idaho, USA. He has authored 9 books and over 450 publications in refereed journals and conferences. His research interests include wireless communications and mobile computing, computer networks, mobile cloud computing, security, and smart grid. He is a Senior Member of ACM. He received teaching awards multiple times from different institutions and the best research award from three institutions. He also served as a member, the chair, and the general chair of a number of international conferences. He was the Chair of the IEEE Communications Society Wireless Technical Committee and the Chair of the TAOS Technical Committee. He served as the IEEE Computer Society Distinguished Speaker from 2003 to 2005. He currently serves on the editorial boards of several international technical journals and is the Founder and the Editor-in-Chief of *Wireless Communications and Mobile Computing* (Wiley). He has guest edited a number of special issues in IEEE journals and magazines.

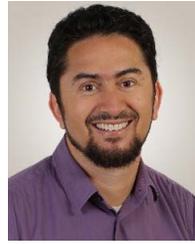

**Bechir Hamdaoui** (S'02–M'05–SM'12) received the Diploma of Graduate Engineer degree from the National School of Engineers at Tunis, Tunisia, in 1997, and M.S. degrees in electrical and computer engineering and computer science and the Ph.D. degree in ECE from the University of Wisconsin-Madison in 2002, 2004, and 2005, respectively. He is currently an Associate Professor with the School of EECS, Oregon State University. His research interest spans various areas in the fields of computer networking, wireless communications, and mobile computing, with a current focus on distributed optimization, parallel computing, cognitive networks, cloud computing, and Internet of Things. He is a Senior Member of the IEEE Computer Society, the IEEE Communications Society, and the IEEE Vehicular Technology Society. He has received several awards, including the ICC 2017 Best Paper Award, the 2016 EECS Outstanding Research Award, and the 2009 NSF CAREER Award. He is currently serving as the Chair for the 2017 IEEE INFOCOM Demo/Posters Program. He has also served as the Chair for the 2011 ACM MOBICOM's SRC Program and the Program Chair/Co-Chair of several IEEE symposia and workshops, including GC 16, ICC 2014, IWCMC 2009–2017, CTS 2012, and PERCOM 2009. He also served on technical program committees of many IEEE/ACM conferences, including INFOCOM, ICC, and GLOBECOM. He has been selected as a Distinguished Lecturer for the IEEE Communication Society for 2016 and 2017. He served as an Associate Editor for the IEEE TRANSACTIONS ON VEHICULAR TECHNOLOGY from 2009 to 2014, *Wireless Communications and Mobile Computing* from 2009 to 2016, and the *Journal of Computer Systems, Networks, and Communications* from 2007 to 2009. He has been an Associate Editor of the IEEE TRANSACTIONS ON WIRELESS COMMUNICATIONS since 2013.

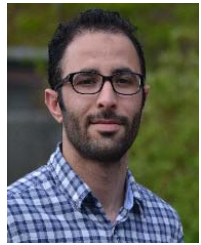

**Bassem Khalfi** (S'14) received the B.S. degree in telecommunications engineering from the l'École supérieure des communications de Tunis (Sup'Com) in 2012, the M.S. degree from the Ecole Nationale d'Ingenieurs de Tunis, Tunisia, and the M.S. degree in mathematics and computer science from Paris Descartes University, France, in 2014. He is currently pursuing the Ph.D. degree with Oregon State University. His research interests include wireless communication and networks, including dynamic spectrum access, wideband spectrum sensing, content centric networking, and IoT.